\documentclass{jpp}
\usepackage{amsmath}
\usepackage{amssymb,amstext}
\usepackage{natbib}
\usepackage{graphicx}

\renewcommand{\vec}[1]{\boldsymbol{#1}}

\newcommand{\diff}[2]{\frac{\mathrm d #1}{\mathrm d #2}}

\title[Compressive high-frequency waves] {Compressive high-frequency waves riding on an Alfv\'en-cyclotron wave in a multi-fluid plasma}

\author[D. Verscharen and E. Marsch]%
  {D\ls A\ls N\ls I\ls E\ls L\ns V\ls E\ls R\ls S\ls C\ls H\ls A\ls R\ls E\ls N
  \thanks{verscharen@mps.mpg.de}
\ns \and\ns
E\ls C\ls K\ls A\ls R\ls T\ns M\ls A\ls R\ls S\ls C\ls H
   \thanks{marsch@mps.mpg.de}}

\affiliation{Max-Planck-Institut f\"ur Sonnensystemforschung, Max-Planck-Stra\ss{}e~2, D-37191~Katlenburg-Lindau, Germany}

\date{3 December 2010, revised 3 February 2011, accepted 4 February 2011}

\begin{document}

\maketitle

\begin{abstract}
We study weakly-compressive high-frequency plasma waves which are superposed
on a large-amplitude Alfv\'en wave in a multi-fluid plasma consisting of
protons, electrons, and alpha particles. For these waves, the plasma
environment is inhomogeneous due to the presence of the low-frequency
Alfv\'en wave with large amplitude, a situation that may apply to space
plasmas such as the solar corona and solar wind. The dispersion relation of
the plasma waves is determined from a linear stability analysis using a new
eigenvalue method that is employed to solve the set of differential wave
equations which describe the propagation of plasma waves along the direction
of the constant component of the Alfv\'en wave magnetic field. This approach
allows one to consider also weak compressive effects. In the presence of the
background Alfv\'en wave, the dispersion branches obtained differ
significantly from the situation of a uniform plasma. Due to compressibility,
acoustic waves are excited and couplings between various modes occur, and
even an instability of the compressive mode. In a kinetic treatment, these
plasma waves would be natural candidates for Landau-resonant wave-particle
interactions, and may thus via their damping lead to particle heating.
\end{abstract}

\section{Introduction}

The solar wind is the classical paradigm of a nonuniform plasma which is
structured in space and time \citep{schwenn91} and permeated by fluctuations
on a wide range of scales. It reveals turbulence and carries large-amplitude
Alfv\'en waves which mostly originate in the solar corona (for reviews of the
magnetohydrodynamic turbulence see, e.g., \citet{tu95, bruno05} and the
plasma kinetics see \citet{marsch06}). The plasma waves in the solar wind are
thus riding on a varying background, and throughout their passage from the
corona the solar wind particles are observed to evolve non-adiabatically, and
therefore are supposed to undergo continuous heating out to 1~AU
\citep[e.g.][]{kasper08} and beyond. It is believed since a long time
\citep{coleman68,belcher71} that wave-particle interactions \citep{marsch06}
with the ubiquitous fluctuations in the electromagnetic fields are the main
driver of this heating process. Here we study weakly-compressive
high-frequency plasma waves which are superposed on a large-amplitude
Alfv\'en wave in a multi-fluid plasma consisting of protons, electrons, and
alpha particles like in the solar wind. The term `high-frequency wave' refers
to the wavenumber regime around the inverse ion inertial length
$\ell_{\mathrm p}$ in contrast to the non-dispersive low-frequency MHD limit.

As is well known, there are two important kinetic resonances that can lead to
dissipative heating of a plasma by wave-particle interactions with waves that
propagate parallel to the background magnetic field. The first is Landau
resonance requiring a parallel wave electric field, and the second cyclotron
resonance \citep[e.g.][]{akhiezer751, hollweg02} that couples to the
perpendicular wave electric field. The condition for Landau resonance is
given by $kv_{\parallel}-\omega=0$, where $k$ denotes the parallel wave
number, $v_{\parallel}$ the particle velocity in the direction parallel to
the background magnetic field and $\omega$ the wave frequency. This effect
can lead to parallel heating of the particles \citep{lehe09}. In a low-beta
plasma such as the solar corona, it is difficult to fulfill this condition if
the wave phase speed is close to the Alfv\'en speed, because
$v_{\mathrm{th}}\ll V_{\mathrm A}$ \citep{chandran10}.

The cyclotron resonance is connected with the transverse electromagnetic
field, and the resonance condition is given by
$kv_{\parallel}-\omega-n\Omega_{j}=0$ with an integer $n$ and the particle
gyrofrequency $\Omega_j$ (only $n=1$ is allowed in the case of parallel
propagation). This resonance can lead to pitch-angle diffusion which is
indeed observed in solar wind protons \citep{heuer07}. The perpendicular
fluctuations must have high wave numbers in the range of the gyroradius to
fulfil the resonance condition. Alfv\'en/ion-cyclotron waves (A/ICs) are
possible candidates for waves that can undergo this kind of interaction.
However, their origin and evolution are not fully understood, even though
their existence in the solar wind was recently proven \citep{jian09}.
Indirect evidence for Alfv\'en-cyclotron heating was already referred to by
several authors beforehand from simulations \citep{gary05} and proton in-situ
observations \citep{marsch01,kasper08}.

The role of weakly compressive waves in the different phenomena of plasma
heating is currently under wide discussion
\citep{tu94,chandran05,bale05,kellogg06,chandran09,verdini10}. Especially,
kinetic Alfv\'en waves (KAWs) have come into the focus of the debate, because
they are both transverse and compressive. However, several problems also
arise from this interpretation of solar wind fluctuations, especially at high
wave numbers \citep{podesta10}.

It has also been known for a long time that compressibility plays a
major role in the context of the parametric instabilities of large-amplitude
waves \citep{galeev63,goldstein78,lashmore-davies79,stenflo07}. These
instabilities are always connected with compressive components of the
daughter-wave products. Our subsequent analysis is based on the previous
derivations by \citet{marsch11}, which treat the density fluctuations in
terms of the longitudinal electrostatic field and a ponderomotive electric
field. The connection between density fluctuations and ponderomotive forces
in the context of parametric instabilities was discussed before by
\citet{sharma83} for many different wave modes. However, the work of those
authors was focused on frequencies around the upper-hybrid frequency, which
is much higher than the frequencies considered here, and thus beyond the
scope of this paper. Electromagnetic circularly polarised waves can also be
generated from the high-frequency side. For example, \citet{murtaza84}
discussed how an upper-hybrid wave can generate such waves in a two-fluid
model. For this purpose, electrostatic effects and ponderomotive forces had
to be included in the model describing the high-frequency pump wave. The
importance of electrons in the decay of compressional Alfv\'en waves was
discussed more recently by \citet{brodin08} in terms of the Hall-MHD
description. These authors found a new decay channel for oblique daughter
waves, and discovered that the wave decay products could grow on scales
around the ion inertial length. They also discussed the role of kinetic
Alfv\'en waves as decay products and their possible ability to heat the
plasma.

Our present work concentrates on purely parallel wave propagation,
but we consider the role of both electrostatic and electromagnetic components
of the wave modes in a multi-fluid plasma. The nonlinear coupling of the
longitudinal electrostatic field and of the ponderomotive electric fields
with the transversal electromagnetic wave fields is the main reason for the
significant changes we found in the mode structure and polarisation. However,
the modified wave modes and the possible decay products are, due to the given
geometry, still forced to propagate along the mean field which is determined
by the constant longitudinal magnetic-field component. So we do not consider
genuine oblique wave propagation.

The multi-fluid wave equations are solved here with an eigenvalue and
eigenvector method that opens a new and unusual way of analysing the pump
wave decay and dispersion properties of the resulting plasma waves. This
approach also provides the dispersion and polarisation properties of these
waves in a comprehensible and direct way.

As mentioned above, it has been known for a long time that the solar wind is
permeated by waves and structures on many different scales \citep{tu95}. But
also in the solar corona low-frequency waves in the magnetic field were
recently observed by remote-sensing techniques \citep{depontieu07}.
Therefore, in theory and modelling it is obviously necessary to assume an
inhomogeneous background magnetic field. The natural choice for such a field
might be one consisting of low-frequency Alfv\'en waves.

\begin{figure}
\vspace*{2mm}
\par
\centering
\includegraphics[width=0.3\textwidth]{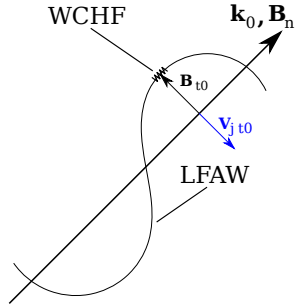}
\caption{Geometry of the scenario. On a constant magnetic field $\vec
B_{\mathrm n}$ along the axis of propagation $\vec k_0$, a low-frequency
Alfv\'en wave (LFAW) propagates and generates the transverse background
fields $\vec B_{\mathrm t0}$ and $\vec V_{j\mathrm t0}=-\zeta\vec B_{\mathrm
t0}$ with $\zeta=\mathrm{const.}>0$. Weakly-compressive high-frequency waves
(WCHF) are superposed in the flank of the Alfv\'en wave.}
\label{fig_situation}
\end{figure}

Consequently, the scenario assumed for our present theoretical treatment is
the following. A low-frequency Alfv\'en-cyclotron wave is assumed to provide
the nonuniform background magnetic field and corresponding background
velocity field, according to the wave polarisation relation. In the flank of
this wave, a linear dispersion and stability analysis is performed for a
three-component plasma consisting of protons, electrons, and alpha particles,
whereby also drifts among these species with respect to each other can be
included. The situation is sketched in Fig.~\ref{fig_situation}. The applied
multi-fluid model allows to consider transverse waves with an intrinsic
weakly compressive component. If transverse dispersion branches at wave
numbers close to the ion gyroradius are found, then these can be made
responsible for possible perpendicular ion heating. Their compressive
electrostatic components can in turn explain parallel heating. Since the
observations indicate large perpendicular temperature anisotropies in the
corona \citep{antonucci00,kohl06} and in fast solar wind \citep{marsch06},
the assumption of weak compressibility seems to be justified empirically, and
is also consistent with the measured density fluctuation level \citep{tu95}.

\section{Theoretical approach and numerical treatment}

\subsection{The multi-fluid model}

Following the detailed derivations of \citet{marsch11}, we obtain
subsequently a set of second-order ordinary coupled differential equations
(DEs) of the relevant fields for parallel propagation, in which case
$B_{\mathrm n}$ is a conserved quantity, and thus the
fields can be made dimensionless as follows: $\vec{v}_{j \mathrm
t}=\vec{V}_{j \mathrm t}/V_{\mathrm A}$, and $\vec{b}_{\mathrm t}=\vec{B}_{\mathrm
t}/B_{\mathrm n}$. Similarly, we define the normalised longitudinal electric
fields as follows:
\begin{equation}
e_{j\mathrm n}=\frac{cE_{j\mathrm n}}{B_{\mathrm n}V_{\mathrm A}}, \; \; \; e_{\mathrm
n}=\frac{cE_{\mathrm n}}{B_{\mathrm n}V_{\mathrm A}}. \label{eq.60}
\end{equation}
Length scales can be all normalised in units of the ion inertial length scale
$\ell_{\mathrm p}=c/\omega_{\mathrm p}$. In the following, all velocities are
normalised to the proton Alfv\'{e}n speed $V_{\mathrm A}=B_{\mathrm
n}/\sqrt{4\pi n_{\mathrm p}m_{\mathrm p}}$ and then denoted by lower case
letters. This choice is different than the normalisation by \citet{marsch11}
but it is more appropriate here, because the wave phase speed $V$ will be
used as a free parameter for different evaluations of the wave equations.
Therefore, it is better to normalise with a fixed velocity. The natural
choice for such a velocity is the proton Alfv\'{e}n speed. In this
normalisation system, frequencies are normalised to the proton gyration
frequency $\Omega_{\mathrm p}$, which is equal to $V_{\mathrm
A}/\ell_{\mathrm p}$. In the subsequent paper, we use $\Omega_j$ as the
normalised gyration frequency of species $j$. We also recall the original
definition
\begin{equation}
e_{j\mathrm n} = \frac{1}{c}\left(\vec v_{j \mathrm t} \times \vec b_{\mathrm
t}\right) \cdot \hat{{\vec n}}. \label{eq.61}
\end{equation}
Using the same normalisation for the transverse electric field, we obtain
\begin{equation}
\vec{e}_{\mathrm t}= - v (\hat{{\vec n}} \times \vec{b}_{\mathrm t}),
\label{eq.62}
\end{equation}
which is fully determined by the solution found for $\vec{b}_{\mathrm t}$,
which means it is a dependent auxiliary field. Completing the required
definitions, we quote the differential operator
\begin{equation}
{\mathcal D}_j = \frac{\mathrm d^2}{\mathrm d \xi^2} + \frac{1}{r_j^2}, \label{eq.63}
\end{equation}
where we have defined the spatial coordinate along $B_\mathrm{n}$ as
$\xi_{\mathrm n}=\xi$, which we will continue using subsequently. After these
preparations we can state the DEs, as derived from the equations for parallel
wave propagation. Firstly, we obtain for each particle species with respect
to its transverse motion a forced harmonic oscillator equation reading
\begin{equation}
{\mathcal D}_j\, \vec{v}_{j \mathrm t} = \left( \frac{e_{\mathrm n}+e_{j\mathrm n}}{c^2_j -
v^2_{j\mathrm n}} \frac{\mathrm d{\vec v}_{j
\mathrm t}}{\mathrm d \xi} + \frac{\vec{b}_{\mathrm t} }{r_j} +
\hat{{\vec n}} \times\frac{\mathrm d \vec{b}_{\mathrm t}}{\mathrm d \xi}
\right) \Omega_j. \label{eq.64}
\end{equation}
Secondly, one can rewrite the mutually coupled and driven wave equations for
the transverse magnetic field,
\begin{equation}
{\mathcal D}_B \, \vec{b}_{\mathrm t} =- \sum_j \frac{1}{\ell_j^2} \left( 
\frac{\vec{v}_{j \mathrm t} }{v_{j\mathrm n}} + \frac{(e_{\mathrm n} +
e_{j\mathrm n}) }{c^2_j -v^2_{j\mathrm n}}(\hat{{\vec n}} \times {\vec v}_{j
\mathrm t}) \right), \label{eq.65}
\end{equation}
and the longitudinal electric field,
\begin{equation}
{\mathcal D}_E \, e_{\mathrm n} = \sum_j \frac{ e_{j\mathrm
n}}{\lambda^2_{j}}. \label{eq.66}
\end{equation}
All involved parameters such as $v_{j\mathrm n}, c_j, \lambda_j, \ell_j$ and
$r_j$ have non-vanishing mean values, as the density $n_j$ is always nonzero
and its fluctuations throughout are assumed to be small. Note that $\Omega_j$
is strictly constant as it depends on the conserved quantity $B_{\mathrm n}$.
The other parameters can be calculated by use of the background number
density $\bar{n}_j$, as well as by exploiting the conditions for
quasi-neutrality, zero longitudinal current, and zero centre-of-momentum
velocity \citep{marsch11}. Compression is accounted for solely by the
longitudinal electric field in this approximation. This is the major
advantage of this multi-fluid system.

Furthermore, the multi-fluid approach incorporates the natural scales of the
plasma and allows therefore the treatment of dispersive waves in the
high-frequency range which is not accessible by MHD considerations. The
compressibility can be accounted for in a comparably lucid way through the
electric field variables $e_{\mathrm n}$ and $e_{j\mathrm n}$. The nonlinear
equations (\ref{eq.64}), (\ref{eq.65}), and (\ref{eq.66}) form a closed set of
DEs, which describe the leading-order compressive effects via $e_{\mathrm
n}(\xi)$. Before we write down these field equations in their components, let
us define an appropriate coordinate system. For a right-handed orthogonal
system, we choose the unit vectors
$\hat{\vec{n}}=\vec{e}_3=\vec{e}_1\times\vec{e}_2$,
$\vec{e}_1=\vec{e}_2\times\hat{\vec{n}}$, and
$\vec{e}_2=\hat{\vec{n}}\times\vec{e}_1$. The normalised (dimensionless)
transverse magnetic field is
\begin{equation}
\vec{b}_{\mathrm t}= b_1(\xi) \vec{e}_1 + b_2(\xi) \vec{e}_2. \label{eq.67}
\end{equation}
Similarly, the transverse flow velocity of any species $j$ is given by
\begin{equation}
\vec{v}_{j \mathrm t}= v_{j1}(\xi)\vec{e}_1 + v_{j2}(\xi)\vec{e}_2.
\label{eq.68}
\end{equation}
The particle speed along the direction of propagation in the fixed coordinate
system (not in the co-moving frame) is denoted by $u_{j\mathrm n}$. If there
are no differential drift motions along the mean magnetic field in the
background plasma, i.e., if for all $j$ we have $u_{j\mathrm n}=0$, then
$v_{j\mathrm n}=-v$, and thus the longitudinal gyration length becomes
$r_j=-v/\Omega_j$, which by definition is not a positive-definite quantity as
the gyrofrequency carries the sign of the charge of the species considered.
Using Eq.~(\ref{eq.67}) and Eq.~(\ref{eq.68}), the longitudinal electric field
associated with species $j$ simply reads
\begin{equation}
e_{j \mathrm n}= v_{j1}b_2 - v_{j2}b_1. \label{eq.69}
\end{equation}
We recall from \citet{marsch11} that $e_{j\mathrm n}$ can also be
expressed as the gradient of a potential, which is given by the transverse
kinetic energy of species $j$ and then has the normalised form,
\begin{equation}
e_{j\mathrm n}= - \frac{1}{\Omega_j} \frac{\mathrm d}{\mathrm d \xi}
\left(\frac{1}{2}\vec{v}_{j \mathrm t}^2\right). \label{eq.70a}
\end{equation}
In terms of components, we have from a comparison of that equation with
Eq.~(\ref{eq.69}) the relation,
\begin{equation}
-\frac{1}{\Omega_j} \frac{\mathrm d v_{j 1,2} }{\mathrm d \xi}= \pm b_{2,1}, \label{eq.70b}
\end{equation}
which we can use to replace the first derivative of the transverse velocity
by the magnetic field. Written out explicitly, the set of DEs in
dimensionless form finally reads as follows:
\begin{align}
\diff{^2v_{j1,2}}{\xi^2}+\frac{v_{j1,2}}{r_j^2}&=\left[ \frac{b_{1,2}}{r_j}\mp\diff{b_{2,1}}{\xi}  \mp\frac{e_n+e_{j\mathrm n}}{c_j^2-v_{j\mathrm n}^2}b_{2,1}\right]\Omega_j \label{eq.71a}\\
\diff{^2b_{1,2}}{\xi^2} -\frac{b_{1,2}}{\ell_{\mathrm S}^2}&=-\sum\limits_j\frac{1}{\ell_j^2}\left[\frac{v_{j1,2}}{v_{j\mathrm n}}\mp \frac{e_n+e_{j\mathrm n}}{c_j^2-v_{j\mathrm n}^2}v_{j2,1}\right] \label{eq.71b} \\
\diff{^2e_{\mathrm n}}{\xi^2}-\frac{e_{\mathrm n}}{\lambda_{\mathrm D}^2}&=\sum\limits_j \frac{1}{\lambda_j^2}(v_{j1}b_2-v_{j2}b_1)  \label{eq.71c}
\end{align}

The operators on the left-hand sides of Eqs.~(\ref{eq.71a}),
(\ref{eq.71b}), and (\ref{eq.71c}) describe the dynamics of the uncoupled free
fields and have a simple physical interpretation. After Fourier
transformation (yielding $\mathrm d/\mathrm d\xi \rightarrow ik$) the
solution of Eq.~(\ref{eq.71a}) gives a helical gyration in space of the
transverse velocity about the mean field $B_{\mathrm n}$ with wave vector
$k=\pm 1/r_j$, the solution of Eq.~(\ref{eq.71b}) corresponds to diamagnetism,
i.e. the static penetration of the transverse field into the plasma by the
skin depth $\ell_{\mathrm S}= c/\omega_{\mathrm P}$, and Eq.~(\ref{eq.71c}) gives
for $v=0$ the static screening by the Debye length $\lambda_{\mathrm D}$, or
for finite speed $v$ the electrostatic wave dispersion relation, $(k
\lambda_{\mathrm D})^2+1=0$. This, for zero drifts, transforms into the
electrostatic dispersion relation
\begin{equation}
k^2 = \sum_j \frac{\omega_j^2}{v^2- c_j^2}, \label{eq.72}
\end{equation}
the zeroes of which which yield the Langmuir and ion-acoustic waves. For
finite right-hand sides of the above DEs, the fields are coupled (note that
the abbreviation from Eq.~(\ref{eq.69}) has to be included). The
incompressible limit (with $e_{\mathrm n}=e_{j {\mathrm n}}=0$) then gives
the monochromatic (only a single $k=k(V)$ is permitted from the dispersion
relation) electromagnetic Alfv\'en-cyclotron wave that has a constant but
arbitrarily large amplitude. When considering compressibility, the electric
and electromagnetic waves are linked and interact through the nonlinear
rightmost terms in the above wave equations. Considering compressibility may
require either a perturbative approach or numerical treatment.

\subsection{Perturbative approach and linearisation}

The above equations are a system of second-order wave equations. To simplify
it, we reduce it to a system of first-order equations. Therefore, we
introduce the quantities $e_{\mathrm n}^{\prime}\equiv\mathrm de_{\mathrm n}/\mathrm d\xi$, $
b_{1,2}^{\prime}\equiv \mathrm d b_{1,2}/\mathrm d\xi$, and
$v_{j1,2}^{\prime}\equiv \mathrm dv_{j1,2}/\mathrm d\xi$. With this
substitution, the system corresponds to a set of 18 coupled nonlinear
first-order ordinary differential equations in $\xi$ for an electron-proton
plasma. For each additional particle species, four coupled equations are
added.

In the next step, the system (\ref{eq.71a}-\ref{eq.71c}) is linearised around
a background given by the wave amplitude vectors $\vec v_{j\mathrm t0}$ and $\vec
b_{j\mathrm t0}$ to determine the wave dispersion. A suitable choice of the
background values is given in Sect.~\ref{sect_background}.

Nonlinear couplings of low-frequency waves with fluctuations at high
frequencies cannot be described by a linearised system. The low-frequency
wave with wave number $k_0$, however, may be treated as a constant
inhomogeneous background if the high-frequency waves with wave numbers $k$
fulfill the condition $k_0\ll k$. The high-frequency waves are then treated
in the flank of this low-frequency wave that does not change its fields
significantly over several periods of the high-frequency waves. The particle
velocities, the electromagnetic field, and their derivatives are combined in
a state vector,
\begin{equation}
\vec y=(v_{\mathrm p1}^{\prime},v_{\mathrm p2}^{\prime},\dots,v_{\mathrm
p1},v_{\mathrm p2},\dots,b_{1}^{\prime},b_{2}^{\prime},b_{1},b_{2},e_{\mathrm
n}^{\prime},e_{\mathrm n}),
\end{equation}
and therewith one can write the linearised equation as,
\begin{equation}\label{ddelydxi}
\frac{\mathrm d}{\mathrm d\xi}\delta \vec y=\mathcal A \delta\vec y,
\end{equation}
with the quadratic matrix $\mathcal A=(a_{i,j})\in M(n)$ with $n=4s+6$ ($s$
is the total number of species). The solution of Eq.~(\ref{ddelydxi}) is in
general given by
\begin{equation}\label{superposition_eigen}
\delta \vec y=\sum \limits_{i=1}^{4s+6}\alpha_i\delta \vec
y_ie^{\lambda_i\xi},
\end{equation}
with the eigenvalues $\lambda_i$ and the corresponding eigenvectors $\delta
\vec y_i$. Since $\mathcal A$ is a real matrix, the complex conjugated
eigenvalues and eigenvectors are also solutions once a complex eigenvalue or
eigenvector is found. The coefficients $\alpha_i$ are arbitrary. However,
they must be equal for the pairwise complex conjugated eigenvalues to
construct a real solution.

An imaginary part of an eigenvalue $\lambda_i$ always indicates a periodic
fluctuation. Due to the rule of pairwise complex conjugated eigenvalues, they
are represented by real sine or cosine functions. Real parts of $\lambda_i$
correspond to growth or damping. We denote the imaginary part as $k$ and the
real part as $\kappa$. If the eigenvalues are pairwise symmetric in the real
part, the growth is described by hyperbolic sine or cosine functions, which
is the symmetric solution for the instabilities growing in positive and in
negative $\xi$-direction.

The eigenvalues can be calculated numerically with the QR-method after
transforming $\mathcal A$ to an upper Hesse matrix \citep{press92}. The
existence of the complex-conjugated eigenvalues means that for wave-like
daughter products always two solutions exist, one foreward and one backward
propagating with the same frequencies.

\subsection{Background wave}\label{sect_background}

An adequate background to calculate the dispersion is the flank of a
circularly polarised Alfv\'en wave, since these waves are exact eigenmodes of
a plasma and obey Eqs.~(\ref{eq.71a}) and (\ref{eq.71b}) in the
incompressible limit with arbitrary amplitudes. The phase speed of this wave
is denoted by $v_0$. It is left-hand polarised and has a magnetic field of
the structure
\begin{equation}
\vec b_{\mathrm t0}=b\begin{pmatrix}\cos(k_0\xi)\\ \sin(k_0\xi)\end{pmatrix}.
\end{equation}
Without loss of generality, we can evaluate this field at the point $\xi=0$.
For sufficient low $k_0$-values, this field appears as a constant magnetic
background field of the magnitude $\vec b_{\mathrm t0}=(b,0)$ as stated
above. Therefore, this background describes analogous conditions as used to
treat oblique propagation of linear modes. However, the wave is additionally
associated with a transversal velocity for each species. Corresponding to
this magnetic field, a background velocity field occurs that is determined by
the polarisation relation of circularly polarised Alfv\'en waves
\begin{equation}
\vec v_{j\mathrm t0}=\frac{v_{j\mathrm
n 0}}{1+kv_{j\mathrm n0}/\Omega_j}\vec b_{\mathrm t0},
\end{equation}
as it is derived by \citet{sonnerup67} for example. The normal velocity
component of the particle species $j$ in the reference frame moving with
$v_0$ is denoted by $v_{j\mathrm n0}$. The polarisation relation is also in
agreement with the results by \citet{marsch11}. For a sufficient small
wavenumber $k_0$, the wave fulfills the Alfv\'enic dispersion relation
$v_0\simeq v_{\mathrm A}$.

The transformation of the transverse velocity to the co-moving reference
frame does not change the value for $\vec v_{j\mathrm t0}$. Therefore, the
polarisation relation provides the necessary (constant) value for $\vec
v_{j\mathrm t0}$ depending on a given (small) $k_0$ and the wave amplitude
$b$.

To evaluate the polarisation relation exactly, the dispersion relation for
the circularly polarised Alfv\'en waves is used. It is given by
\begin{equation}
k_0^2+ \sum \limits_{j=1}^{s}\frac{\omega_j^2}{c^2}
\frac{k_0v_{j\mathrm n0}}{k_0v_{j\mathrm n0}+\Omega_j}=0
\end{equation}
as the non-relativistic limit from \citet{sonnerup67}.

\section{Dispersion of high-frequency waves}

In this section we study the wave dispersion relation of high-frequency waves
propagating on the nonuniform background Alfv\'en wave. The most important
free parameter of the system of equations (\ref{eq.71a}-\ref{eq.71c}) is the
disposable wave phase speed $v$. For a fixed $v$, the eigenvalue/eigenvector
method provides the corresponding wavenumber values $k(v)$. The relation
$\omega=kv$ then delivers the corresponding $\omega$, and thus by scanning
through all relevant values of $v$ the full dispersion relation can be
determined. The growth rate can be determined similarly by evaluating
$\gamma=\kappa v$ with the spatial growth rate $\kappa$ from the eigenvalue
determination. The very high-frequency branches, which are dominated by the
electron dynamics, are not treated in detail, since the ions carry the main
momentum, and their wave-induced motion is therefore more important for the
heating and acceleration processes. The electron density and relative
velocity with respect to the protons and the other ionic species is
determined by the requirement of vanishing charge density (quasineutrality)
and vanishing constant longitudinal currents of the background, such that
\begin{align}
\sum \limits_{j=1}^{s}n_jq_j&=0, \label{chargedens}\\
\sum \limits_{j=1}^sn_jq_jv_{j\mathrm n}&=0.
\end{align}
The specific-heat ratio is set to its adiabatic value of $\gamma_j=5/3$ for
all species.

\subsection{Homogeneous background}

\begin{figure}
\vspace*{2mm}
\par
\centering
\includegraphics[width=\textwidth]{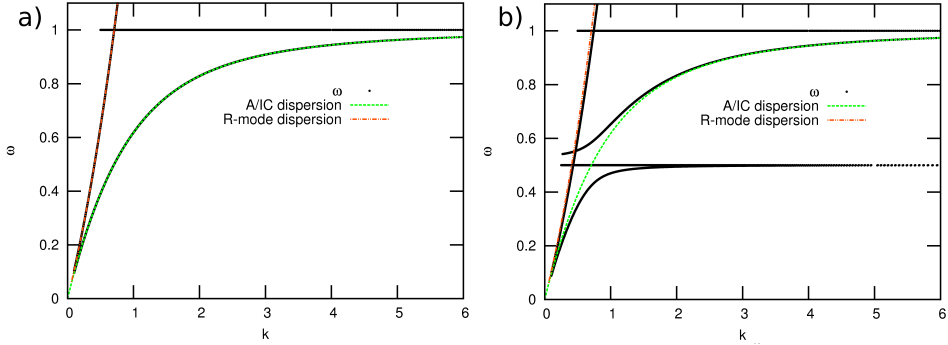}
\caption{a) Dispersion relation for an electron-proton plasma in a
homogeneous background with $\beta=0$. The cold dispersion relation for
Alfv\'en/ion-cyclotron waves and for ion-corrected R-mode waves are
additionally plotted. b) The same for a plasma consisting of electrons,
protons, and alpha particles with $\beta=0.01$.} \label{fig_mf2_aic}
\end{figure}

Before we discuss the case of an inhomogeneous plasma, we apply the system of
equations and the linearisation to an electron-proton plasma for a
homogeneous background and with $\beta_j=0$. The plasma beta is defined as
the ratio of the thermal energy density to the magnetic field energy density
for each species. The background field is set to $B_{\mathrm n}=5 \times
10^{-5}\,\mathrm G$ and the particle number densities to $n_{\mathrm
p}=n_{\mathrm e}=5\,\mathrm{cm}^{-3}$. The ratio $B_{\mathrm n}^2/n_{\mathrm
p}$ must be parameterised even in the dimensionless normalisation of the
system, in order to fix the normalised plasma frequency
$\omega_{j}/\Omega_{\mathrm p}$ for all electrostatic processes. The result
of the calculation is shown in Fig.~\ref{fig_mf2_aic}a. The two
incompressible branches correspond to the transverse particle motions. They
represent a free motion with the gyration frequency for the protons at
$\omega=1$, their normalised gyrofrequency. The branch approaching the
gyrofrequency asymptotically corresponds to the Alfv\'en/ion-cyclotron wave.
In Fig.~\ref{fig_mf2_aic}, the cold dispersion relation for
Alfv\'en/ion-cyclotron waves
\begin{equation}
\left(\frac{\omega}{\Omega_{\mathrm p}}\right)^2
=\ell_{\mathrm p}^2k^2+\frac{\ell_{\mathrm p}^4k^4}{2}-\frac{\ell_{\mathrm p}^3k^3}{2}\sqrt{\ell_{\mathrm p}^2k^2+4}
\end{equation}
is additionally plotted as green dotted line. The branch at low $k$-values
with high phase speed $v$ corresponds to the whistler mode shown as red
dashed line. The corrected low-frequency dispersion for the multi-fluid
R-mode wave (i.e., the whistler wave with ionic effects) is given
analytically by
\begin{equation}
\omega=\frac{\Omega_{\mathrm e}}{2\left(1+\frac{1}{\ell_{\mathrm
e}^2k^2}\right)}\left[\sqrt{1+\frac{4}{\ell_{\mathrm p}^2k^2}}+1\right].
\end{equation}
Both relations show a perfect agreement with the numerical calculations made
with the eigenvalue method \citep{stix92}. This confirms the validity of our
approach.

In Fig.~\ref{fig_mf2_aic}b, the dispersion relation is shown for a plasma
consisting of electrons, protons, and alpha particles without any relative
drifts and without a background wave field. The particle species have a beta
of 0.01. The proton number density is set to $n_{\mathrm
p}=5\,\mathrm{cm}^{-3}$, the alpha particle number density to
$n_{\alpha}=0.04n_{\mathrm p}$ and the electron number density according to
Eq.~(\ref{chargedens}). The alpha-particle cyclotron branch approaches
asymptotically the frequency $\omega=1/2$ as expected. This value corresponds
to the alpha particle gyrofrequency in normalised units. The cyclotron branch
of the protons is slightly deformed in the low-$k$ range in comparison to the
cold plasma dispersion relation. Also the whistler wave mode is slightly
shifted. The presence of the alpha particles is responsible for this
deviation. Further correction terms would be needed to represent the
dispersion relation in this three-fluid plasma analytically.

\begin{figure}
\vspace*{2mm}
\par
\centering
\includegraphics[width=\textwidth]{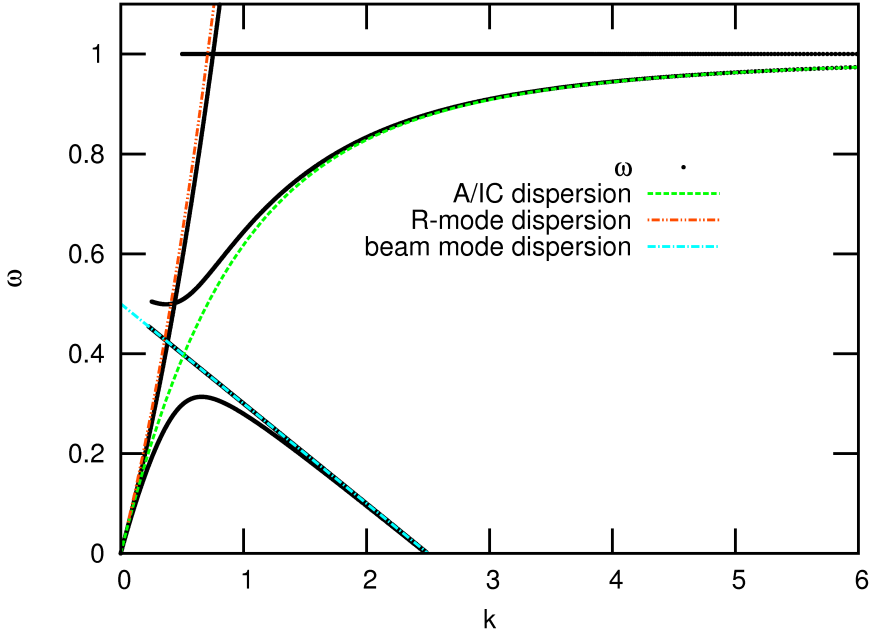}
\caption{Dispersion relation for a plasma consisting of protons, electrons,
and drifting alpha particles. The beam mode occurs for drifting alpha
particles.} \label{fig_mf2_drift}
\end{figure}

Our model also allows to include relative drifts of the particles along the
wave normal direction. The relative drift speed between protons and alpha
particles is defined as $v_{\mathrm d}\equiv v_{\mathrm{pn}}-v_{\alpha
\mathrm n}$. In Fig.~\ref{fig_mf2_drift}, the dispersion relation for
$v_{\mathrm d}=0.2$ is shown. The alpha-particle branch starting at
$\omega=\Omega_{\alpha}$ is turned into the so-called beam-mode branch (line
inclined to the left) derived from the resonance condition
\begin{equation} \omega=\Omega_{\mathrm
\alpha}+ku_{\alpha \mathrm n},
\end{equation}
where $u_{\alpha \mathrm n}=v-v_{\alpha \mathrm n}$ denotes again the
alpha-particle bulk speed component in the wave normal direction in the
proton rest frame. For a vanishing drift $u_{\mathrm n \alpha}$, this mode is
flattened back to the horizontal line $\omega=\Omega_{\alpha}=1/2$.

\subsection{Inhomogeneous background}

\begin{figure}
\vspace*{2mm}
\par
\centering
\includegraphics[width=\textwidth]{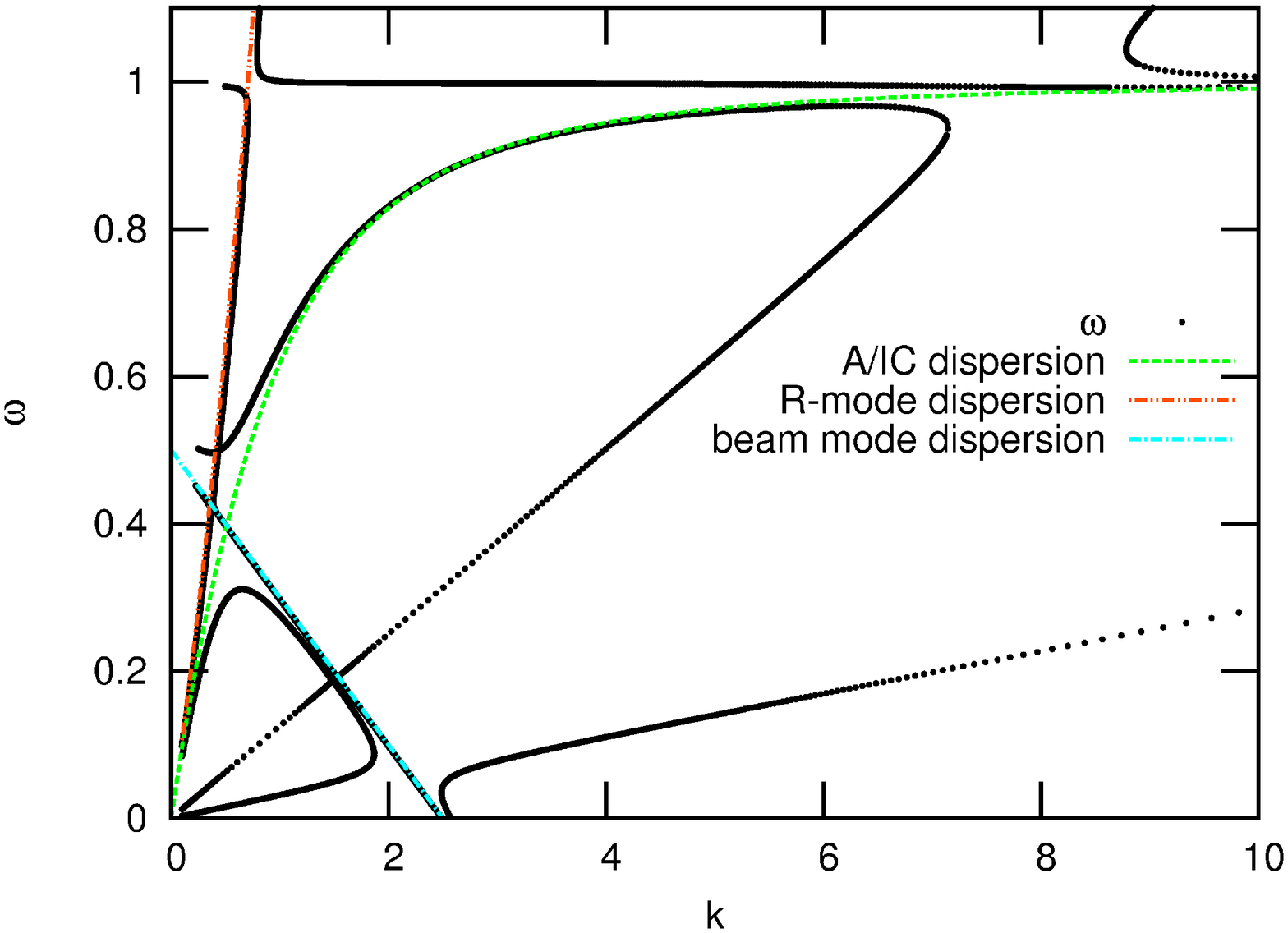}
\caption{Dispersion relation for a plasma consisting of protons, electrons,
and drifting alpha particles. The frequency $\omega$ is shown in dependence
on the wave number $k$ in dimensionless units. The derived A/IC branch
deviates from the cold dispersion branch at low and high wave numbers. Mode
coupling between the various branches occurs as discussed in the text.}
\label{fig_mf2_sw}
\end{figure}

Next, our system of coupled wave equations can be applied to an inhomogeneous
background plasma, which corresponds to realistic solar wind conditions. The
plasma consists of three species: protons, electrons, and alpha particles.
The relative drift between protons and alphas is set to $v_{\mathrm
d}=v_{p\mathrm n}-v_{\alpha \mathrm n}=0.2$, and the plasma beta to
$\beta_j=0.01$ for each species. The background wave is assumed to have a
normalised wave number of $k_0=0.01$ and an amplitude of $b=0.1$. Such a wave
has a phase speed of almost the local proton Alfv\'en speed. The wave number
is small enough to allow us to neglect any direct nonlinear couplings between
the background wave and the high-frequency waves. For the latter, the
background wave appears as a quasi-constant field with respect to which the
system can be linearised. The results are shown in Fig.~\ref{fig_mf2_sw}.

The dispersion branches are deformed for several reasons. The proton
Alfv\'en/ion-cyclotron branch at lower wave numbers is not further deformed
as compared to Fig.~\ref{fig_mf2_aic}b. Some branches turn at some
bifurcation positions into completely different dispersion branches. These
transitions correspond to mode couplings. For example, the faster sound wave
(alpha-particle mode) couples with the Alfv\'en/ion-cyclotron wave of the
protons at high wavenumbers, whereas the ion-acoustic wave couples with the
drift-deformed Alfv\'en/ion-cyclotron wave of the alpha particles. The R-mode
couples with the ion gyration at $\omega=1$ but, interestingly enough, not
with the alpha-particle beam mode. It shows a break at $\omega=1$ and then
continues above that frequency.

Two additional linear branches occur with merely constant phase speeds of
$v\approx 0.13$ and $v\approx 0.03$. Such linear acoustic modes as these two
were also found by \citet{mann97} for a warm plasma. The wave phase speed of
these modes is usually determined by
\begin{align}
v_{\mathrm{Ph}1}&=\frac{\gamma_{\mathrm p}k_{\mathrm B}T_{\mathrm p}
+ \gamma_{\mathrm e}k_{\mathrm B}T_{\mathrm e} }{m_{\mathrm p}}\\
v_{\mathrm{Ph}2}&=\frac{\gamma_{\alpha}k_{\mathrm B}T_{\alpha}}{m_{\alpha}}
\end{align}
for a plasma with $n_{\alpha} \ll n_{\mathrm p}$. The first is the so-called
ion-acoustic speed, the second is the sound speed of the alpha-particle
component. In our case, these velocities are given by $v_{\mathrm{Ph}1}=0.13$
and $v_{\mathrm{Ph}2}=0.23$. The first one corresponds perfectly with the
fast (steeper line) sound wave that we have found. The drift velocity
$v_{\mathrm d}$ of the alpha particles is the reason for the deviation of the
second linear mode. In the non-drifting reference case, the two velocities
match ($v_{\mathrm{Ph}2}-v_{\mathrm d}=0.03$). If the numerical dispersion
code is applied to a plasma without drift, the second sound wave branch is
directly found at $v=v_{\mathrm{Ph}2}$ (not shown here).

The two sound wave modes do not appear in the homogeneous plasma. Remember,
that the phase speed $v$ is a free parameter in our calculation, and a
wavenumber is given by the eigenvalues obtained for each $v$. If a wave is
not dispersive, and therefore has a constant phase speed, it is not possible
to calculate the full set of possible $k$-values that belong to this $v$. The
sound waves exist already in the homogenous case; however, they become
visible only after becoming dispersive due to the mode coupling. This type of
wave becomes only dispersive for non-zero beta and in the presence of the
background wave. Also above the corresponding gyrofrequencies $\omega=1$ and
$\omega=1/2$, the branches continue with their constant phase speed.

The waves on all branches have a compressive component, owing to the
non-vanishing electric fields $e_{\mathrm n}$ and $e_{j\mathrm n}$. The real
part of the eigenvalues $\lambda_i$ is always zero, i.e., none of the
compressive modes is unstable in this case.

Note that it is the amplitude of the background wave which mainly determines
the strength of the mode coupling and, therefore, the position and shape of
the deformation of the ion-cyclotron branch. For higher amplitudes, the
branch turns earlier away from the A/IC dispersion branch and, hence, the
phase speed of the linear mode increases already at lower $k$-values. The
plasma beta determines the phase speeds of the linear branches. Higher betas
lead to higher phase speeds of these modes. The phase speeds can be adjusted
relative to each other by choosing different betas for the individual
species. However, the overall topology of the dispersion branches is not
significantly changed by assigning different beta values to the different
species.

\begin{figure}
\vspace*{2mm}
\par
\centering
\includegraphics[width=\textwidth]{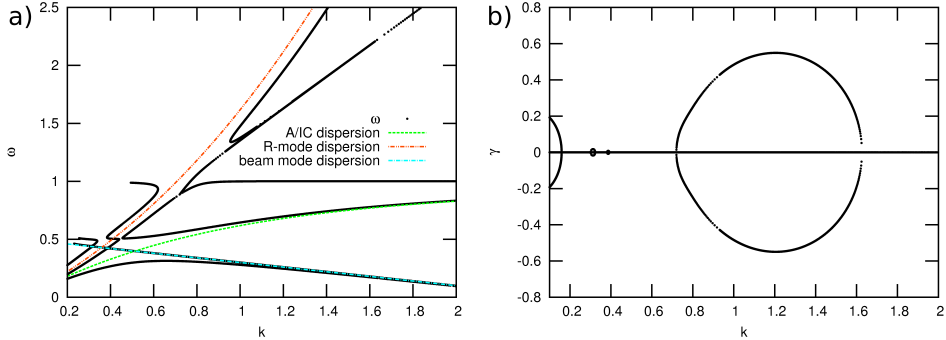}
\caption{a) Dispersion relation for a plasma with $\beta=1.2$. b) Imaginary
part of the frequency depending on $k$.} \label{fig_mf2_low}
\end{figure}

However, if beta is chosen to be greater than one, the situation changes
tremendously. The sound waves become unstable. In Fig.~\ref{fig_mf2_low}a,
the dispersion is shown for a situation with $\beta_j=1.2$. The growth rate
of the modes is shown in Fig.~\ref{fig_mf2_low}b.

Mode coupling between the sound wave and the whistler wave mode leads to an
instability that occurs at $k\approx0.8\dots 1.6$. Interestingly enough, this
range is the ion gyroradius scale for moderate plasma betas close to one.
This growing instability can therefore provide wave energy to the dissipative
regime by means of Landau resonance, since the particle distribution has a
significant number of particles at the corresponding wave speed of about
$v\approx 1.36$ for a plasma with $\beta_j \gtrsim 1$. This instability
vanishes if the background wave is absent, i.e., in the homogeneous case. For
higher beta values, the position of the instability is shifted to higher
wavenumbers.

\section{Conclusions}

As expected, the inhomogeneity of the background wave field leads to a
deformation of the standard normal modes in the plasma. All of the discovered
waves have a compressive component. Compressibility and inhomogeneity of the
background plasma are the reasons for the new effects in the dispersion
properties. The initially non-compressive Alfv\'en/ion-cyclotron waves become
slightly compressive due to the inhomogeneous background, and are therefore
able to interact with the protons via their electrostatic field components,
in addition to the cyclotron resonance enabled by their transverse
components. The quantitative details of these wave-particle interactions
cannot be treated in the fluid description, since they are purely kinetic
processes. Therefore they require a kinetic Vlasov treatment at least in the
frame of quasi-linear theory and are, therefore, beyond the scope of this
work. The kinetic refinement of the dispersion analysis of circularly
polarised waves of the same type was described by \citet{stenflo76}, who also
showed how relativistic effects and compressibility can modify the dispersion
of large-amplitude waves in multi-fluid theory.

The non-constant (due to the presence of the pump wave) background
leads to nonlinear mode couplings between some of the linear wave modes. It
also leads to the excitation of initially non-dispersive modes, such as the
ion-acoustic wave or the alpha-particle sound wave, yet now with
$k$-dependent wave speeds.

The acoustic modes can grow at wavenumbers around the ion gyration scale
under certain conditions and for adequate parameters. Thus, they are good
candidates for a longitudinal electrostatic wave field with which particles
can undergo Landau-resonant wave-particle interactions. Kinetic Alfv\'en
waves (KAWs) have recently been discussed as another possible reason for this
kind of interaction. KAWs owe their compressibility to the oblique geometry
of their propagation with respect to a constant background field. In this
study, we can show that also waves with purely parallel propagation can grow
nonlinearly at the corresponding resonant wavenumbers, provided that the
background has a non-trivial---but quite reasonable---magnetic field
configuration. These two mechanisms are quite different in nature and should
be further investigated and compared with each other.

The instability of ion-acoustic modes coupled to transverse modes were also
discussed in studies of the parametric decay of large-amplitude pump waves
and evaluated by numerical simulations \citep{araneda07,valentini09}. In
these studies, also instabilities of longitudinal waves at high wavenumbers
are found, however at different background parameters. These waves are
usually interpreted as results of nonlinear wave-wave interactions. Our
linearised wave equations, however, can explain a similar growth of daughter
waves at higher frequencies than the initial pump wave frequency as a
consequence of compressibility and a non-uniform background. This mechanism
can thus be understood as a new spectral transfer process of plasma
fluctuations. It is very similar to the decay instability, which is found
with the characteristic $k\gg k_0$.

Our eigenvector analysis still keeps the freedom to choose the amplitudes
$\alpha_i$ in Eq.~(\ref{superposition_eigen}). This means that a wave can in
general only occur if its amplitude is finite. It is beyond the scope of this
work to investigate how the discovered wave branches can be excited in a real
plasma. The linear dispersion analysis can only show possible normal modes.
The discovered instability leads to growth in the initial phase only until
nonlinear couplings and perhaps saturation occur. Yet unstable modes can grow
from the thermal noise (that is constrained as a finite eigenvector of the
system of equations) with a certain finite amplitude. In the presence of a
large-amplitude wave, this thermal noise can lead to growth according to the
calculated growth rate $\gamma$ at the gyroradius scale range for $\beta
\gtrsim 1$. Maybe, full nonlinear calculations can investigate the further
evolution and possible nonlinear excitations of such modes. The linear
approximation appears to reflect the basic situation well \citep{lehe09}.

In the future, also other background conditions should be inquired. The above
chosen background is one of the simplest inhomogeneous conditions. It is
important to note that in our approach the background has to change slowly in
dependence upon the position $\xi$. Otherwise, the use of a fixed phase speed
$v$ is not possible anymore, and then the present approximation needs to be
changed. The original set of coupled wave equations remains valid, yet
another mathematical treatment is required to cope with their nonlinearity.

\begin{acknowledgment}
D.~V. is grateful for financial support by the International Max Planck Research
School (IMPRS) on Physical Processes in the Solar System and Beyond.
\end{acknowledgment}

\bibliographystyle{jpp}
\bibliography{weakly_compressive_paper}

\end{document}